\documentclass[twocolumn]{aastex63}

\newcommand\herschel{{\itshape Herschel}}

\newcommand\spitzer{{\itshape Spitzer}}
\newcommand\wise{{\itshape WISE}}

\received{---}
\revised{---}
\accepted{---}

\shorttitle{Far-infrared excess from the west hot spot of Pictor A}
\shortauthors{Isobe et al.}

\begin{document} 
\title{\herschel\ SPIRE discovery of far-infrared excess synchrotron emission 
from the west hot spot of the radio galaxy Pictor A.}

\author{Naoki Isobe}
\affiliation{
	Institute of Space and Astronautical Science (ISAS), 
        Japan Aerospace Exploration Agency (JAXA), 
        3-1-1 Yoshinodai, Chuo-ku, Sagamihara, Kanagawa, 252-5210, Japan}
\email{n-isobe@ir.isas.jaxa.jp}
\author{Yuji Sunada}
\affiliation{
	Department of Physics, Saitama University, 
        255 Shimo-Okubo, Sakura-ku, Saitama, 338-8570, Japan}
\author{Motoki Kino}
\affiliation{
	Kogakuin University of Technology \& Engineering, 
	Academic Support Center,
	2665-1 Nakano, Hachioji, Tokyo, 192-0015, Japan}
\affiliation{
	National Astronomical Observatory of Japan,
       2-21-1 Osawa, Mitaka, Tokyo, 181-8588, Japan}
\author{Shoko Koyama}
\affiliation{Academia Sinica Institute of Astronomy and Astrophysics,
	PO Box 23-141, Taipei 10617, Taiwan}
\author{Makoto Tashiro}
\affiliation{
	Institute of Space and Astronautical Science (ISAS), 
        Japan Aerospace Exploration Agency (JAXA), 
        3-1-1 Yoshinodai, Chuo-ku, Sagamihara, Kanagawa, 252-5210, Japan}
\affiliation{
	Department of Physics, Saitama University, 
        255 Shimo-Okubo, Sakura-ku, Saitama, 338-8570, Japan}
\author{Hiroshi Nagai}
\affiliation{
	National Astronomical Observatory of Japan,
       2-21-1 Osawa, Mitaka, Tokyo, 181-8588, Japan}
\affiliation{ 
	The Graduate University for Advanced Studies (SOKENDAI), 
	Osawa 2-21-1, Mitaka, Tokyo, 181-8588, Japan}
\author{Chris Pearson}
\affiliation{ 
	RAL Space, STFC Rutherford Appleton Laboratory, 
	Didcot, Oxfordshire, OX11 0QX, UK}
\affiliation{ 
	Oxford Astrophysics, University of Oxford, 
	Keble Rd, Oxford OX1 3RH, UK}
\affiliation{ 
	Department of Physical Sciences, 
	The Open University, Milton Keynes, MK7 6AA, UK}

\begin{abstract}
A far-infrared counterpart to the west hot spot of the radio galaxy Pictor A 
is discovered with the Spectral and Photometric Imaging REceiver (SPIRE)
onboard \herschel.
The color-corrected flux density of the source is measured 
as $70.0 \pm 9.9$ mJy at the wavelength of 350 \micron.
A close investigation into its radio-to-optical spectrum 
indicates that the mid-infrared excess over the radio synchrotron component,
detected with \wise\ and \spitzer, significantly contributes to the far-infrared band.
Thanks to the SPIRE data, it is revealed that 
the spectrum of the excess is described by a broken power-law model
subjected to a high-energy cutoff. 
By applying the radiative cooling break 
under continuous energy injection ($\Delta \alpha = 0.5$),
the broken power-law model supports an idea  
that the excess originates in 10-pc scale substructures within the hot spot.
From the break frequency,  $\nu_{\rm b} = 1.6_{-1.0}^{+3.0} \times 10^{12}$ Hz, 
the magnetic field was estimated as $B\simeq1$--$4$ mG. 
This is higher than the minimum-energy magnetic field of the substructures
by a factor of $3$--$10$.
Even if the origin of the excess is larger than $\sim 100$ pc,
the magnetic field stronger than the minimum-energy field is confirmed.
It is proposed that regions with a magnetic field locally boosted 
via plasma turbulence are observed as the substructures. 
The derived energy index below the break, 
$\alpha \sim 0.22$ (conservatively $<0.42$), 
is difficult to be attributed to the strong-shock acceleration ($\alpha = 0.5$). 
Stochastic acceleration and magnetic reconnection are considered 
as a plausible alternative mechanism.
\end{abstract}

\section{Introduction} 
\label{sec:intro}      
Compact radio hot spots hosted by Fanaroff-Riley type-II \citep[FR-II;][]{fanaroff74} 
radio galaxies are widely associated with a strong shock 
at the terminal of their nuclear jets \citep[e.g.,][]{begelman84}.
The existence of high-energy electrons in the hot spots,
accelerated by the shock via the first-order Fermi process,
is observationally evidenced  by their bright synchrotron radio emission 
and inverse-Compton X-ray one \citep[e.g.,][]{hardcastle04}.
These make the hot spots one of the possible generators of high-energy cosmic 
rays \citep[e.g.,][]{meisenheimer89, meisenheimer97}. 
For detailed investigation of the acceleration phenomena in the hot spots, 
it is necessary to put observational constraints on the physical parameters,
such as the magnetic field strength,
by making use of the multi-frequency spectrum. 

The radio and X-ray spectral energy distributions of the hot spots are 
typically described by a simple one-zone synchrotron-self-Compton (SSC) model. 
The magnetic field in the hot spots estimated by the SSC model is usually 
consistent to the minimum-energy and/or equipartition value 
\citep[e.g.,][]{hardcastle04,kataoka05}.
However, the spectrum of low-luminosity objects tends 
to deviate from the one-zone SSC model.
By pioneering mid-infrared (MIR) studies with \spitzer,
a hint of a multi-synchrotron scenario was given 
to several low-luminosity hot spots  \citep[e.g.,][]{kraft07,werner12},
and also to some jet knots \citep[e.g.,][]{uchiyama06}.
Thus, the acceleration phenomena in the hot spots 
are suggested to be more complicated and inhomogeneous 
in comparison to the standard picture of the strong shock acceleration.

The prototypical FR II radio galaxy Pictor A  \citep{perley97},
located at the redshift of $z = 0.035$ \citep{eracleous04},
provides an ideal laboratory 
for an observational study of the particle acceleration in the hot spots.
The west hot spot of this radio galaxy is one of the brightest and 
most extensively studied low-luminosity hot spots 
in a variety of frequency ranges from the radio to X-ray bands
\citep[e.g.,][]{thomson95,meisenheimer97,tingay08,wilson01,werner12}.
Because the X-ray spectrum of the object is 
reported to be unaccountable by the one-zone SSC model,
a number of the multi-synchrotron ideas were 
invoked \citep[e.g.,][]{wilson01,tingay08}. 

In addition, a new spectral component was identified 
from the west hot spot of Pictor A in the MIR band \citep{isobe17}. 
From a careful analysis of the MIR data
obtained with the Wide-field Infrared Survey Explorer (\wise),
a significant MIR excess was unveiled 
over the power-law (PL) like extrapolation of the synchrotron radio spectrum
smoothly connecting to the optical one. 
By adopting a naive cutoff-PL model to the excess, 
a number of important implications are derived by \citet{isobe17}.
Firstly, they proposed that the MIR excess originates in the 10-pc scale substructures
within the hot spot resolved by the Very Long Baseline Array \citep[VLBA,][]{tingay08}.
Secondly, the magnetic field in the substructures are possible to be amplified
via some plasma effects 
in comparison to the minimum-energy field.
Thirdly, they hinted that the spectrum of the excess 
(the energy index of $\alpha \sim 0.4$)
is harder than the prediction of the strong shock acceleration ($\alpha = 0.5$).
However, the detailed investigation into the spectrum of the excess 
has remained yet unexplored,
due to the wavelength gap between the radio and MIR bands. 

In order to precisely specify the physical condition 
for the electron acceleration of the excess,
it is indispensable to determine precisely its spectral shape. 
The \herschel\ space observatory \citep{pilbratt10} 
has a potential to tackle this issue, 
since it is able to widely cover the far-infrared (FIR) range 
with a wavelength of $\lambda = 60 $--$670$ \micron,
corresponding to the frequency of $\nu = (0.45$--$5)\times 10^{12}$ Hz. 
It is found that the west hot spot of Pictor A was serendipitously mapped 
with the Spectral and Photometric Imaging REceiver 
\citep[SPIRE;][]{griffin10} onboard \herschel\ at its three photometric bands 
with an effective wavelength of $\lambda= 250$, $350$ and $500$ \micron.
On these SPIRE maps, 
an FIR source is discovered at the position of the hot spot. 
The SPIRE data, in combination with the previous multi-wavelength ones,
make possible a detailed re-examination on the spectrum of the excess and 
on its physical interpretation. 

In the following, the cosmological constants 
same as those in \citet{tingay08} and \citet{isobe17} are employed
for mutual consistency; 
$H_{\rm 0} = 71$ km s$^{-1}$ Mpc$^{-1}$, 
$\Omega_{\rm m} = 0.27$, and  $\Omega_{\Lambda} = 0.73$ \citep{spergel03}. 
Under this cosmology, an angular size of  $1$ \arcsec\ 
corresponds to a physical size of $688$ pc at the rest frame of Pictor A 
($z = 0.035$).

\section{FIR data with the \herschel\ SPIRE}
\subsection{Observation and Analysis}   
\label{sec:observation}
The nucleus of the FR-II radio galaxy Pictor A was previously 
targeted by the \herschel\ space observatory on three occasions. 
In one observation conducted with the SPIRE photometer in the small-map mode 
on 2011 September 22 (Obs. ID = 1342229235),
the west hot spot of Pictor A was serendipitously mapped. 
In the remaining two observations, 
in which the Photodetector Array Camera and Spectrometer \citep[PACS;][]{poglitsch10}
was operated in the photometer scan-map mode, 
the object was unfortunately located outside the PACS field of view.

The present study focuses on the SPIRE observation of Pictor A. 
The final version of the SPIRE science data were obtained 
from the \herschel\ Science Archive. 
The Levels 1 and 2 products of the SPIRE data were analyzed  
with the version 15.0.1 of the \herschel\ Interactive Processing Environment (HIPE)
by referring to the latest version of the calibration tree ({\tt spire\_cal\_14\_3}). 

In order to carry out detection and position determination of FIR point sources 
on the SPIRE map,
the HIPE tool SUSSEXtractor ({\tt sourceExtractorSussextractor}) was adopted. 
Photometry of the detected sources was performed with 
the SPIRE Timeline Fitter ({\tt sourceExtractorTimeline}) in the standard manner.
This method is reported to be most reliable 
among the four major photometric procedures implemented by the HIPE
for a point source brighter than $\sim 30$ mJy \citep{pearson14}. 
The source coordinates determined with the SUSSEXtractor were 
employed as a reference position for the SPIRE Timeline Fitter.
It was confirmed that the photometric results 
with the SPIRE Timeline Fitter and SUSSEXtractor
become consistent to each other within a difference of $\lesssim 10$\%,
which is smaller than the photometric
error at all the three SPIRE bands.

\begin{figure*}[t!]
\plottwo{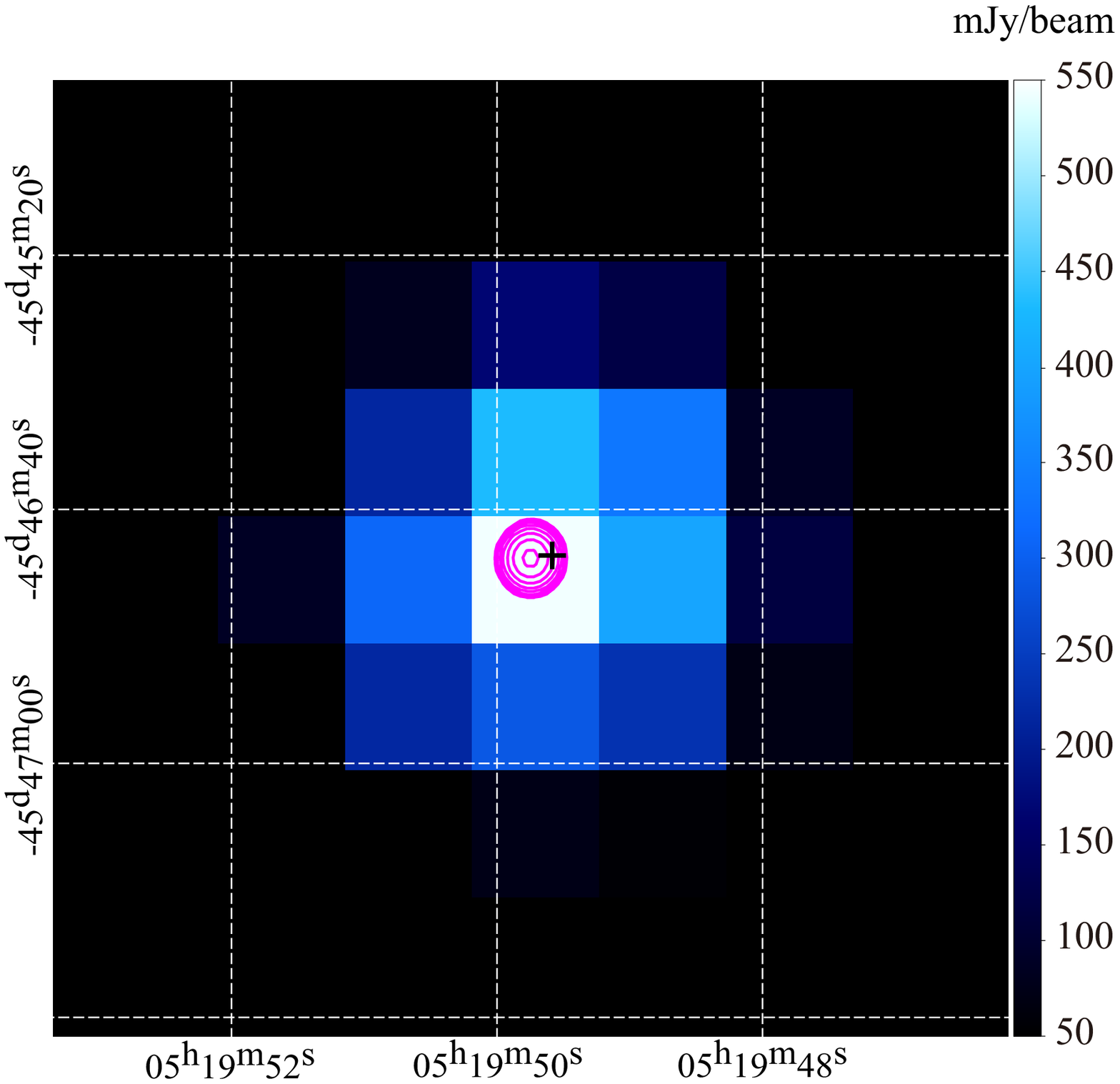}{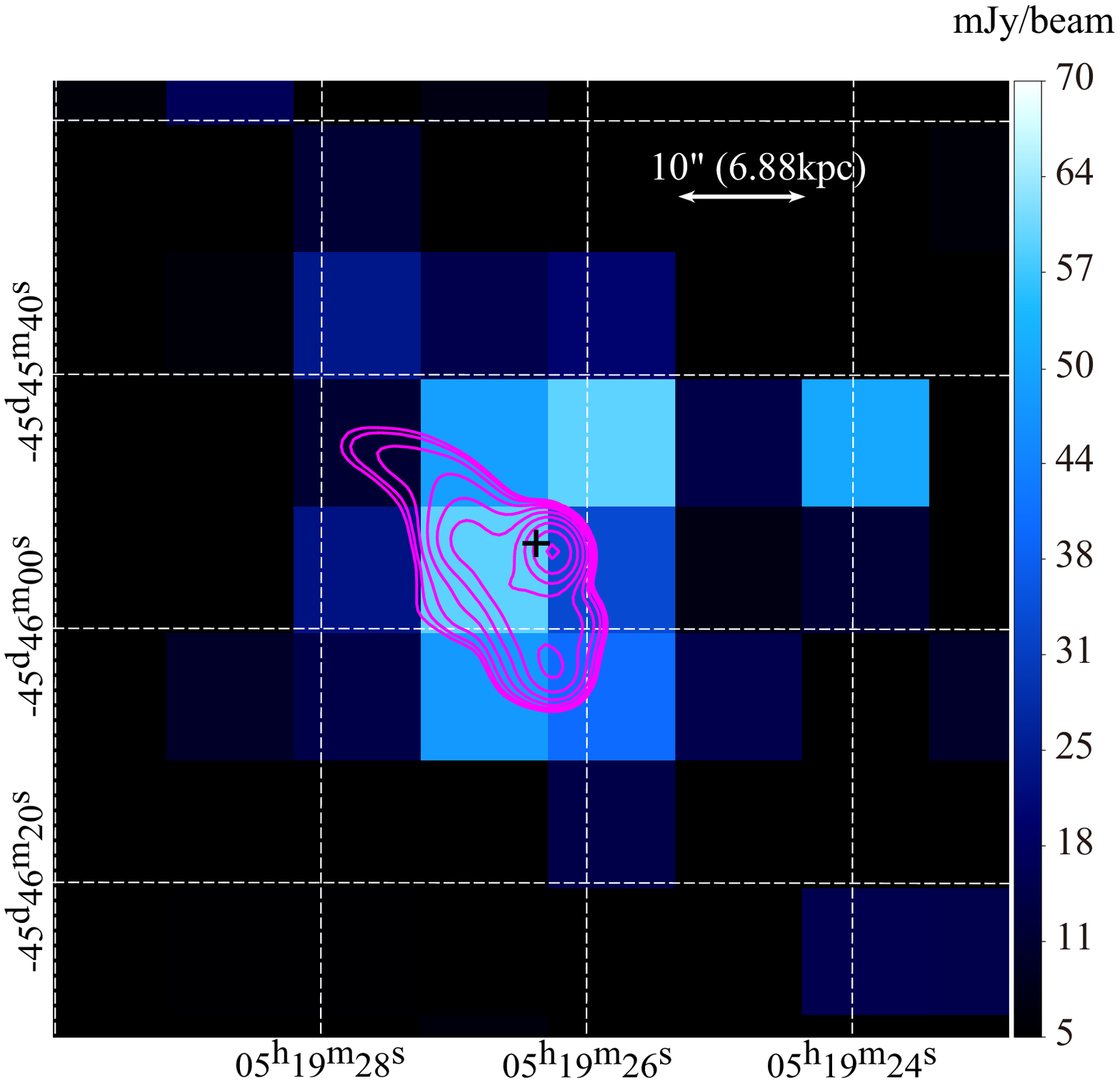}
\caption{
SPIRE Level-2 point-source maps at $350$ \micron\ 
around the nucleus (left) and west hot spot (right) of 
the radio galaxy Pictor A.
The color scale in the individual panels 
indicates the FIR intensity in the unit of mJy beam$^{-1}$,
where the $350$ \micron\ SPIRE beam size is 
$24.2\arcsec$ in the full width at half maximum.  
The 4.8 GHz ATCA image is overlaid with contours.
The major and minor radii of the ATCA beam are evaluated 
as $1.17 \arcsec$ and $1.25 \arcsec$, respectively.
The cross in the individual panels shows 
the $350$ \micron\ SPIRE position of the nucleus and hot spot. 
The horizontal arrow in the right panel corresponds 
to the angular size of $10\arcsec$ ($6.88$ kpc at the source rest frame). }
\label{fig:image}
\end{figure*}

\subsection{Result}   
\label{sec:result}
The SPIRE Level-2 point-source maps ({\tt psrcPMW})
around the nucleus and west hot spot of Pictor A are compared 
in the left and right panels of Figure \ref{fig:image}, respectively,  
with the radio image obtained with the Australia Telescope Compact Array (ATCA)
at the 4.8 GHz band 
\citep[][the original data were kindly provided by Dr. E. Lenc, private communication]{isobe17}. 
Although the spatial resolution is highest at $250$ \micron\ 
among the three SPIRE photometric bands,
the significance of the FIR source corresponding to the west hot spot 
is relatively low at this band (see below).
Thus, the SPIRE images at the wavelength of $350$ \micron\ are utilized here.

The left panel of Figure \ref{fig:image} clearly displays that 
FIR emission is detected from the nucleus
with a $350$ \micron\ color-uncorrected flux density of $618 \pm 7$ mJy. 
The SPIRE position of the nucleus at $350$ \micron,  
indicated with the cross in the left panel of Figure \ref{fig:image},
coincides well with its 4.8 GHz ATCA peak 
within the SPIRE astrometric uncertainty 
\citep[$\sim2 \arcsec$;][]{swinyard10}. 

An FIR source spatially associated with the west hot spot of Pictor A 
is clearly found on the 350 \micron\ SPIRE image
shown in the right panel of Figure \ref{fig:image}.
This FIR source is significantly detected at all the three SPIRE photometric bands 
with a signal-to-noise ratio of  $4.6$, $7.1$ and $10.7$ at $250$, $350$ and $500$ \micron, 
respectively. 
The sky position of the SPIRE source, 
indicated by the cross in the right panel of Figure \ref{fig:image}, 
is found to show a reasonable coincidence to the ATCA position of the west hot spot,
within the SPIRE astrometric accuracy.  
Thus, in the following, 
the SPIRE source is regarded as the real FIR counterpart of the west hot spot. 

Table \ref{tab:photometry} summarizes the SPIRE photometry of the west hot spot
performed with the Timeline Fitter. 
The color-corrected FIR flux density of the object
was derived as $F_{\nu} = 70.0 \pm 9.9$ mJy  at $350$ \micron.
The color-correction factor $f_{\rm col}$ was taken 
from the SPIRE calibration tree 
by assuming the energy index of $\alpha = 1.1$ ($F_\nu \propto \nu^{-\alpha}$).
The adopted index was found to be self-consistent to the color-corrected SPIRE spectrum 
of which the energy index was measured as $\alpha = 1.13 \pm 0.04$. 
At least in the range of $\alpha = 0 - 2$, 
the impact of the color correction is negligible 
(i.e., $f_{\rm col} = 0.99$--$1.00$) for all the three SPIRE bands.

\begin{deluxetable}{lccc}[htb]
\tablecaption{SPIRE photometry of the west hot spot of Pictor A.
\label{tab:photometry}}
\tablecolumns{4}
\tablewidth{0pt}
\tablehead{
	\colhead{$\lambda$ ($\mu$m) \tablenotemark{a} } & 
	\colhead{$F_\nu$ (mJy) \tablenotemark{b} } &
	\colhead{$\Delta F_{\rm \nu,sys}$ (mJy) \tablenotemark{c} } &
       \colhead{$f_{\rm col}$ \tablenotemark{d} }
	}
\startdata 
	 $250$	& $63.3 \pm 10.7$	& $3.5$ & $ 1.0003$ \\
	 $350$	& $70.0 \pm 9.9$	& $3.9$ & $ 1.0004$ \\
	 $500$	& $123.7 \pm 10.9 $	& $6.8$ & $ 0.9995$\\
\enddata
\tablenotetext{a}{The effective wavelength.}
\tablenotetext{b}{The color-corrected flux density.}
\tablenotetext{c}{The systematic error of the SPIRE photometry 
                  \citep[$5.5$\%;][]{bendo13,pearson14}.}
\tablenotetext{d}{The adopted color-correction factor.} 
\end{deluxetable}

\begin{figure*}[htbp!]
\plotone{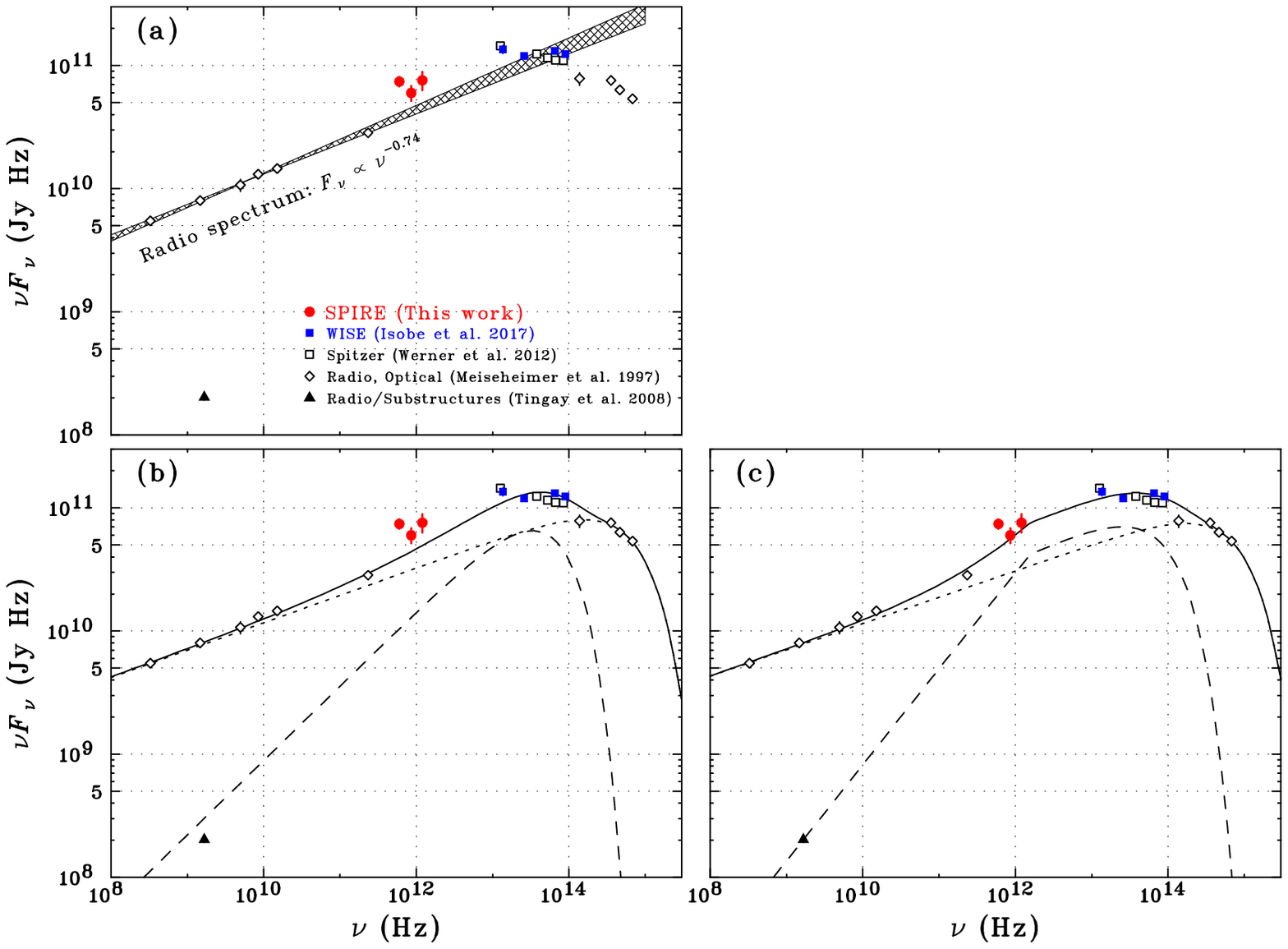}
\caption{Synchrotron spectral energy distribution 
of the west hot spot of Pictor A in the radio-to-optical range. 
The filled circles indicate the FIR data taken with the \herschel\ SPIRE, 
while the filled and open boxes show the MIR data 
with \wise\ \citep{isobe17} and \spitzer\ \citep{werner12}, respectively.
The radio and optical data \citep{meisenheimer97} 
are plotted with the open diamonds.
The filled triangle indicates the $1.67$ GHz VLBA flux 
summed over the substructures within the west hot spot \citep{tingay08}.
In panel (a), the best-fit PL model to the radio data from 330 MHz to 233 GHz
is shown with the hatched tie. 
The solid line in panel (b) corresponds to the double cutoff-PL model 
by \citet{isobe17}, which is composed of the main (dotted line) and 
excess (dashed line) components. 
Panel (c) visualizes that the composite model
(Case 2 in Table \ref{tab:SED_fit}),  
where the broken-PL model is adopted for the excess
with the radio flux fixed at the total flux of the substructures 
(i.e., 121.3 mJy, the filled triangle),
successfully reproduced the radio-to-optical spectrum.}
\label{fig:SED}
\end{figure*}

\section{Spectral energy distribution} 
\label{sec:SED}
\subsection{FIR excess} 
\label{sec:excess}
Figure \ref{fig:SED} compiles the spectral energy distribution of the synchrotron radiation 
from the west hot spot of Pictor A in the radio-to-optical frequency range 
\citep[][]{meisenheimer97,wilson01,tingay08,werner12,isobe17}. 
The best-fit PL model to the radio data in the 330 MHz--233 GHz range
with an energy index of $\alpha = 0.74 \pm 0.015$ \citep{meisenheimer97} 
is shown with the hatched tie in panel (a) of Figure \ref{fig:SED}.
The PL-like radio spectrum appears to be smoothly connected 
to the optical data points, with a cutoff at $\nu \gtrsim 10^{14}$ Hz. 
However,  \citet{isobe17} unveiled that 
the MIR flux measured with \wise\ and \spitzer\ \citep{werner12} significantly exceeds 
the extrapolation of the radio spectrum  
especially at $\nu \lesssim 3 \times 10^{13}$ Hz (i.e., $\lambda \gtrsim 10$ \micron).
They tried to reproduce the radio-to-optical synchrotron spectrum of the object 
by a double cutoff-PL model, as shown in panel (b) of Figure \ref{fig:SED}.
The main cutoff-PL component (the dotted line in panel b of Figure \ref{fig:SED})
dominates the flux in the radio and optical bands, 
while the second cutoff-PL one (the dashed line) 
was introduced to reproduce the MIR excess. 

Panel (a) of Figure \ref{fig:SED} clearly displays that 
the FIR flux obtained with the \herschel\ SPIRE (the filled circles)
is significantly higher than the PL model fitted to the radio data. 
This strongly indicates that the excess has a significant contribution 
not only in the MIR range but also in the FIR one. 
As shown in panel (b), the double cutoff-PL model by \citet{isobe17}
is still inadequate to describe the FIR data. 
Therefore, the spectral shape of the FIR-to-MIR excess 
requires to be re-modeled. 
The FIR excess flux in the SPIRE range is summarized in Table \ref{tab:excess}.
By adopting the model parameters presented in \citet{isobe17},
the excess flux density above the main component is evaluated 
as $F_{\rm \nu, ex1} = 33.4\pm 9.9$ mJy at the wavelength of 350 \micron,
while that above the double cutoff-PL model 
is estimated as $F_{\rm \nu, ex2} =18.4\pm 9.9$ mJy. 
It is impossible to ascribe the FIR excess 
to the systematic uncertainty in the SPIRE photometry,
$\Delta F_{\rm \nu, sys}=3.9$ mJy at 350 \micron\ 
\citep[typically $5.5$\%;][]{bendo13,pearson14}. 

\begin{deluxetable}{lccc}[htb]
\tablecaption{The FIR excess flux measured with the SPIRE.  \label{tab:excess}}
\tablecolumns{4}
\tablewidth{0pt}
\tablehead{
	\colhead{$\lambda$ (\micron)} & 
	\colhead{250} &
	\colhead{350} &
       \colhead{500} 
	}
\startdata
$F_{\nu, \rm{ex1}}$ (mJy) \tablenotemark{a}
	& $35.2\pm 10.7$	& $33.4\pm 9.9$	&$75.4\pm 10.9 $\\
$F_{\nu, \rm{ex2}}$ (mJy) \tablenotemark{b}
	& $22.1\pm 10.7$	& $18.4\pm 9.9$	&$58.1\pm 10.9 $\\
\enddata
\tablenotetext{a}{The excess flux over the main cutoff-PL component of \citet{isobe17}.}
\tablenotetext{b}{The excess flux over the total model of \citet{isobe17}.}
\end{deluxetable}

\subsection{Rough evaluation on the spectrum of the excess} 
\label{sec:2point}
It is proposed by \citet{isobe17} that the excess component originates 
in the 10-pc scale substructures within the west hot spot 
resolved by the VLBA \citep{tingay08}, 
while the main component is attributed to the integrated emission all over the object. 
Based on this two-zone scenario, the spectrum of the excess is
re-evaluated below. 

In order to roughly grasp the spectral shape, 
a simple two-point spectral index between the frequencies $\nu_1$ and $\nu_2$,
defined as 
\begin{equation}
\alpha_{\nu_1-\nu_2} = -\frac{\log F_\nu({\nu_1}) - \log F_\nu({\nu_2})}
                    {\log \nu_1 - \log \nu_2}, 
\end{equation}
is generally useful. 
First, the two-point index between the FIR and MIR bands is evaluated. 
Here, the excess flux over the main component 
(i.e., $F_{\nu, \rm{ex1}}$ in Table \ref{tab:excess}) is adopted. 
With the spectral parameters taken from \citet{isobe17}, 
the flux density of the excess in the \wise\ range is calculated 
as $F_{\nu, \rm{ex1}} = 5.79 \pm 0.76$  mJy 
at $\nu = 1.36 \times 10^{13}$ Hz ($\lambda = 22$ \micron). 
A comparison of the SPIRE and \wise\ flux yields 
the FIR-to-MIR two-point spectral index of the excess 
as $\alpha_{\rm F-M} \simeq 0.6$--$0.8$.

Next, the low-frequency index below the SPIRE frequency range is estimated. 
As the radio intensity of the excess, 
the sum of the VLBA flux density of the substructures,
calculated as $F_{\nu} = 121.3 \pm 5.8$ mJy at $1.67$ GHz after \citet{tingay08},
is adopted. 
This flux density is indicated with the filled triangle in Figure \ref{fig:SED}. 
By connecting the SPIRE and VLBA data, 
the spectral slope between the radio and FIR bands is evaluated 
as $\alpha_{\rm R-F} \simeq 0.1$--$0.2$.

A comparison of the two indices, $\alpha_{\rm R-F}$ and $\alpha_{\rm F-M}$, 
strongly suggests that the excess exhibits a spectral break 
around the SPIRE frequency range.  
It is important to note that the difference between the two indices,
$\Delta\alpha = \alpha_{\rm{F-M}} - \alpha_{\rm{R-F}} \simeq  0.5$,
seems consistent to the radiative cooling break   
under the continuous energy injection \citep[e.g.,][]{carilli91}.

\begin{deluxetable*}{lllll}[htb]
\tablecaption{Spectral parameters derived by the two-component model.
\label{tab:SED_fit}}
\tablecolumns{5}
\tablewidth{0pt}
\tablehead{
        \colhead{Component} & 
        \colhead{Model}     & 
        \colhead{Parameter} & 
        \colhead{Case 1} &
        \colhead{Case 2} 
	}
\startdata
Main 	& cutoff-PL & $F_{\nu}$(5 GHz) (Jy) \tablenotemark{a}	  
			& $2.03 \pm 0.12$  	
			& $1.98 \pm 0.09$\\
	&           & $\alpha$ \tablenotemark{b}                  
			& $0.78 \pm 0.02$ 		
			& $0.79 \pm 0.02$ \\
     	&           & $\nu_{\rm c}$ ($10^{14}$ Hz) \tablenotemark{c}       	
			& $7.1^{+3.5}_{-2.3} $ 	 
			& $ 8.0^{+3.9}_{-2.6} $\\ 	
\hline 
Excess	& broken-PL & $F_{\nu}$(1.67 GHz) (mJy) \tablenotemark{d} 
			& $55^{+327}_{-47}$
			& $121.3$ (fix)\\ 
	&           & $\alpha_{\rm low}$ \tablenotemark{e}	  
			& $0.10 \pm 0.32$ 
			& $0.22 \pm 0.06 $\\
	&           & $\nu_{\rm b}$ ($10^{12}$ Hz) \tablenotemark{f}	  
			& $0.9^{+7.2}_{-0.8}$  
			& $1.6^{+3.0}_{-1.0} $ \\
	&           & $\nu_{\rm c}$ ($10^{13}$ Hz) \tablenotemark{c}	  
			& $6.7^{+6.3}_{-3.3}$ 
			& $ 8.8^{+3.4}_{-2.4} $ \\
\enddata
\tablenotetext{a}{The 5 GHz flux density.}
\tablenotetext{b}{The energy index of the cutoff PL model.}
\tablenotetext{c}{The cutoff frequency.}
\tablenotetext{d}{The 1.67 GHz flux density.}
\tablenotetext{e}{The spectral index below the break frequency.
                  The index above the break is assumed to be $\alpha_{\rm high} = \alpha_{\rm low}+0.5$.}
\tablenotetext{f}{The break frequency.}
\end{deluxetable*}

\subsection{Spectral Modeling} 
\label{sec:SEDmodel}
Based on the argument of the two-point index, 
the cutoff-PL model to the excess adopted in \citet{isobe17} is replaced 
with a broken-PL model subjected to a high-frequency cutoff
(hereafter, simply refereed to as the broken-PL model). 
The cooling break condition under the continuous injection,
i.e. $\Delta\alpha =  0.5$ at the break, is applied to the broken-PL model.

The SPIRE data, in combination with the radio, MIR and optical one,
are examined by the two-component model 
consisting of the main cutoff-PL and excess broken-PL components. 
With the normalization of the excess component left free, 
a reasonable solution is obtained by the composite model.
The resultant spectral parameters are listed 
in Table \ref{tab:SED_fit} (Case 1). 
All the parameters of the main component 
stay unchanged within errors from those in \citet{isobe17}.
The model yields the 1.67 GHz flux density and 
energy index below the break of the excess component 
as $F_{\nu} =55^{+327}_{-47}$ mJy and $\alpha_{\rm low}=0.10 \pm 0.32$, 
respectively.
The derived flux density agrees with 
the sum flux density of the VLBA substructures (i.e., $121.3$ mJy at 1.67 GHz).
This result supports the idea that 
the FIR-to-MIR excess is produced in the substructures.

By fixing the flux density of the excess component 
at the total flux density  of the substructures, 
the composite model successfully describes the spectral energy distribution 
of the west hot spot, as shown in panel (c) of Figure \ref{fig:SED}
where the main cutoff-PL and excess broken-PL components are drawn with 
the dotted and dashed lines, respectively. 
Table \ref{tab:SED_fit} tabulates the derived spectral parameters (Case 2).  
It is suggested that 
both components exhibit a comparable flux in the FIR-to-MIR frequency range 
(namely in $10^{12}$--$5\times10^{13}$ Hz), 
while the excess has only a minor contribution below the SPIRE frequency range.
The break frequency of the excess is measured as 
$\nu_{\rm b} = 1.6_{-1.0}^{+3.0} \times 10^{12}$ Hz. 
The low-frequency index below this break, 
determined as $\alpha_{\rm low} = 0.22 \pm 0.06$, 
becomes consistent to the two-point index, 
$\alpha_{\rm{R-F}} \simeq 0.1$--$0.2$. 
The cutoff frequency of the excess,
$\nu_{\rm c} = 8.8_{-2.4}^{+3.4} \times 10^{13}$ Hz, 
is an order of magnitude lower than that of the main component,
$\nu_{\rm c} = 8.0^{+3.9}_{-2.6} \times 10^{14}$ Hz. 
In comparison to the the cutoff-PL modeling by \citet{isobe17}, 
the cutoff frequency of the excess becomes slightly higher.

\section{Discussion}      
\label{sec:discussion}
\subsection{Overview of the SPIRE results} 
\label{sec:summary}
On the serendipitous maps obtained with the \herschel\ SPIRE photometer,
the FIR counterpart of the west hot spot of the radio galaxy Pictor A 
has been discovered with a color-corrected $350$ \micron\ flux density 
of $F_{\nu} = 70.0 \pm 9.9$ mJy. 
A careful investigation of the radio-to-optical synchrotron spectral energy distribution
shown in Figure \ref{fig:SED} 
reveals that the MIR excess detected with \wise\ and \spitzer\ \citep{isobe17}
significantly extends into the FIR range.

The spectral energy distribution of the object between the radio and optical bands 
is re-analyzed by the two-zone model, comprised of the main and excess components.
Owing to the SPIRE data, it is found that 
the spectrum of the excess is better reproduced by the broken-PL model
rather than the simple cutoff-PL one, 
while the spectral shape of the main component 
agrees with the result of \citet{isobe17} within the errors. 
On the condition of the radiative cooling 
under the continuous energy injection 
\citep[i.e., $\Delta \alpha = 0.5$;][]{carilli91}, 
the broken-PL model becomes consistent to the scenario that 
the 10-pc scale substructures within the hot spot 
detected with the VLBA \citep{tingay08}, 
of which the integrated flux is $121.3$ mJy,
is the generator of the FIR-to-MIR excess.
Based on this idea, 
the break frequency and low-frequency index of the excess is determined  
as $\nu_{\rm b} = 1.6_{-1.0}^{+3.0} \times 10^{12}$ Hz 
and $\alpha_{\rm low} = 0.22 \pm 0.06$, respectively (Case 2). 
In the following, by making the most of the derived spectral parameters,
the physical condition to produce the FIR-to-MIR excess is discussed.

The shortest baseline of $l=236$ km between Pie Town and Los Alamos 
adopted in the VLBA observation by \citet{tingay08}
corresponds to an angular scale of $\arctan(\lambda/l) = 0.16"$ 
($\sim 100$ pc at the redshift of Pictor A)
at the wavelength of $\lambda = 18$ cm (i.e, $1.67$ GHz). 
As a result, this VLBA observation is likely insensitive to 
a spatial structure roughly larger than a few 100 pc.
Thus, the 10-pc scale substructures may represent 
only a tip of a larger spatial structure.
For a quantitative evaluation on the interpretation 
that the 100-pc scale region is the source of the excess,
the following constraints on the radio flux and energy index
are adopted from the spectral modeling (Case 1);
$F_{\nu} < 382$ mJy at 1.67 GHz and $\alpha_{\rm low} < 0.42$,
respectively.

\subsection{Magnetic field estimates} 
\subsubsection{Minimum energy condition} 
\label{sec:Bme}
The minimum-energy condition \citep[e.g.,][]{miley80}
is widely adopted to quantify the magnetic field strength 
in synchrotron-emitting sources,
especially when their inverse-Compton emission is not yet observed. 
The spectral properties of well-studied hot spots 
are reported to be successfully explained 
by the minimum-energy magnetic field $B_{\rm me}$  
\citep[e.g.,][]{hardcastle04,kataoka05}.

Here, the $B_{\rm me}$ value of the substructures are evaluated,
since they are regarded as the origin of the FIR-to-MIR excess.  
The radio properties of the individual substructures taken from \citet{tingay08} 
are tabulated in Table \ref{tab:Bme}.
The VLBA image indicates that 
the minor axis of the substructures roughly corresponds 
to the direction of the jet
from the Pictor A nucleus toward the west hot spot \citep{tingay08}. 
Thus, the spatial geometry of the substructures is assumed to be a simple disk,
and their major and minor lengths are adopted for the diameter and thickness of the disk, 
respectively. 

The method presented in \citet{miley80} is utilized 
to calculate the minimum-energy magnetic field.
For simplicity, a PL-like synchrotron spectrum with an index of $\alpha_{\rm low} = 0.22$
is adopted for all the substructures.
The lower end of the synchrotron frequency is set at $\nu_{\rm min} = 10$ MHz 
by referring to \citet{miley80}.
The hardness of the spectrum ensures that  
the $B_{\rm me}$ estimate is insensitive to $\nu_{\rm min}$. 
The energy of the synchrotron emission is integrated 
up to the break frequency $\nu_{\rm b}$.
Because of the spectral break of $\Delta \alpha =0.5$, 
the contribution of the synchrotron energy above $\nu_{\rm b}$ is 
regarded as negligible.
The proton energy is neglected for the evaluation 
\citep[i.e., the proton-to-electron energy ratio of $k=0$ in][]{miley80}.

As shown in Table \ref{tab:Bme}, 
the minimum-energy magnetic field is estimated as $B_{\rm me}=0.25$--$0.43$ mG. 
This is found to be comparable to that of the whole west hot spot,
$B_{\rm me}=0.31$ mG, which is re-evaluated for $k=0$ from \citet[][]{isobe17}
by assuming a sphere geometry with a diameter of $500$ pc \citep{wilson01}. 
The derived $B_{\rm me}$ values for the substructures and entire west hot spot
are plotted against the thickness $T$ of the region in Figure \ref{fig:B-R}
(here and hereafter, the diameter is regarded 
as the thickness for the sphere geometry).

As briefly mentioned in \S\ref{sec:summary},
the FIR-to-MIR excess is possible to be attributed 
to a region larger than $\sim 100$ pc,
although such a region is potentially resolved out 
in the VLBA observation of \citet{tingay08}.
In order to check this interpretation, 
an upper limit on $B_{\rm me}$ is roughly estimated 
by utilizing those on the flux density and energy index of the excess,
$F_{\nu} = 382$ mJy at 1.67 GHz and  $\alpha_{\rm low} = 0.42$ respectively.
A sphere geometry is simply adopted for the evaluation. 
The dashed line in Figure \ref{fig:B-R} shows the $B_{\rm me}$ upper limit 
as a function of the source thickness.
At $T=200$ pc, 
the constraint on the minimum-energy magnetic field is given 
as $B_{\rm me}<0.32$ mG.

\begin{deluxetable*}{lcccc}[htb]
\tablecaption{Estimates of the minimum-energy magnetic field. 
\label{tab:Bme}}
\tablecolumns{3}
\tablewidth{0pt}
\tablehead{
	\colhead{Component}	& \colhead{$F_{\nu}$(1.67 GHz) (mJy) \tablenotemark{a}}
				& \colhead{Diameter (pc) \tablenotemark{a}}	
				& \colhead{Thickness (pc) \tablenotemark{a}} 	
				& \colhead{ $B_{\rm me}$ (mG) } 
	}
\startdata
Substructure A \tablenotemark{b}	
	& $28.3 \pm 2.8$ & $170$	& $42 $	& $0.31$	 \\
Substructure B \tablenotemark{b}	
	& $25.9 \pm 2.6$ & $87 $	& $46 $	& $0.43$	 \\
Substructure C \tablenotemark{b}	
	& $24.9 \pm 2.5$ & $127$	& $43 $	& $0.35$	 \\
Substructure D \tablenotemark{b}	
	& $ 6.0 \pm 0.6$ & $138$	& $28 $	& $0.25$	 \\
Substructure E \tablenotemark{b}	
	& $36.2 \pm 3.6$ & $122$	& $58 $	& $0.36$	\\
West hot spot \tablenotemark{c}	
	& ---			& ---		&---		& $0.31$  \\
\enddata
\tablenotetext{a}{Taken from \citet{tingay08}.}
\tablenotetext{b}{A disk geometry is adopted.}
\tablenotetext{c}{A sphere geometry with a diameter of 500 pc \citep{wilson01} is adopted.}
\end{deluxetable*}

\subsubsection{Cooling break} 
\label{sec:B_cooling}
A more reliable estimate of the magnetic field is derived 
by assuming that the spectral break is caused by the radiative cooling. 
At the cooling break, the electron radiative cooling timescale 
$t_{\rm rad} = \frac{ 3 m_{\rm e} c}{ 4 u_{\rm B} \sigma_{\rm T} \gamma}$
becomes equal to the adiabatic loss timescale $t_{\rm ad} = \frac{T} {v}$,
where $\gamma$ is the Lorentz factor of synchrotron electrons, 
$m_{\rm e}$ is the electron rest mass, 
$u_{\rm B}=\frac{B^2}{8\pi}$ is the  energy density of the magnetic field $B$,
$\sigma_{\rm T}$ is the Thomson cross section, $T$ is the thickness of the source,
$v$ is the downstream flow velocity in the shock frame,
and $c$ is the speed of light \citep[e.g.,][]{inoue96}. 
Here, the effect of the inverse-Compton cooling is neglected 
based on the previous studies on the west hot spot of Pictor A
\citep[e.g.,][and references therein]{tingay08, isobe17}.
Thus, the electron Lorentz factor at the cooling break is given as   
$\gamma_{\rm b} = \frac{6 \pi m_{\rm e} v c }{\sigma_{\rm T} B^{2} T}$. 

The cooling break was related to the cutoff frequency of 
the cutoff-PL model to the excess in \citet{isobe17}. 
However, the SPIRE data reveals that the broken-PL model,
satisfying the cooling break condition under the continuous energy injection 
($\Delta \alpha = 0.5$),
is more appropriate to reproduce the excess.
Therefore, 
the break frequency of the broken-PL model is actually regarded as  
the cooling break of the excess (i.e., $\nu_{\rm b} \propto \gamma_{\rm b}^2$).
In this case, the break frequency is converted into the magnetic field as follows; 
\begin{equation}
B^3 \simeq 
\frac{27 \pi e m_{\rm e} v^2  c}{\sigma_{\rm T}^2} T^{-2} \nu_{\rm b}^{-1},
\label{eq:Bme}
\end{equation}
where $e$ denotes the elementary charge.
Instead, the cutoff frequency $\nu_{\rm c}$ is thought to reflect 
the maximum Lorentz factor of the accelerated electrons. 

\begin{figure}[htbp!]
\plotone{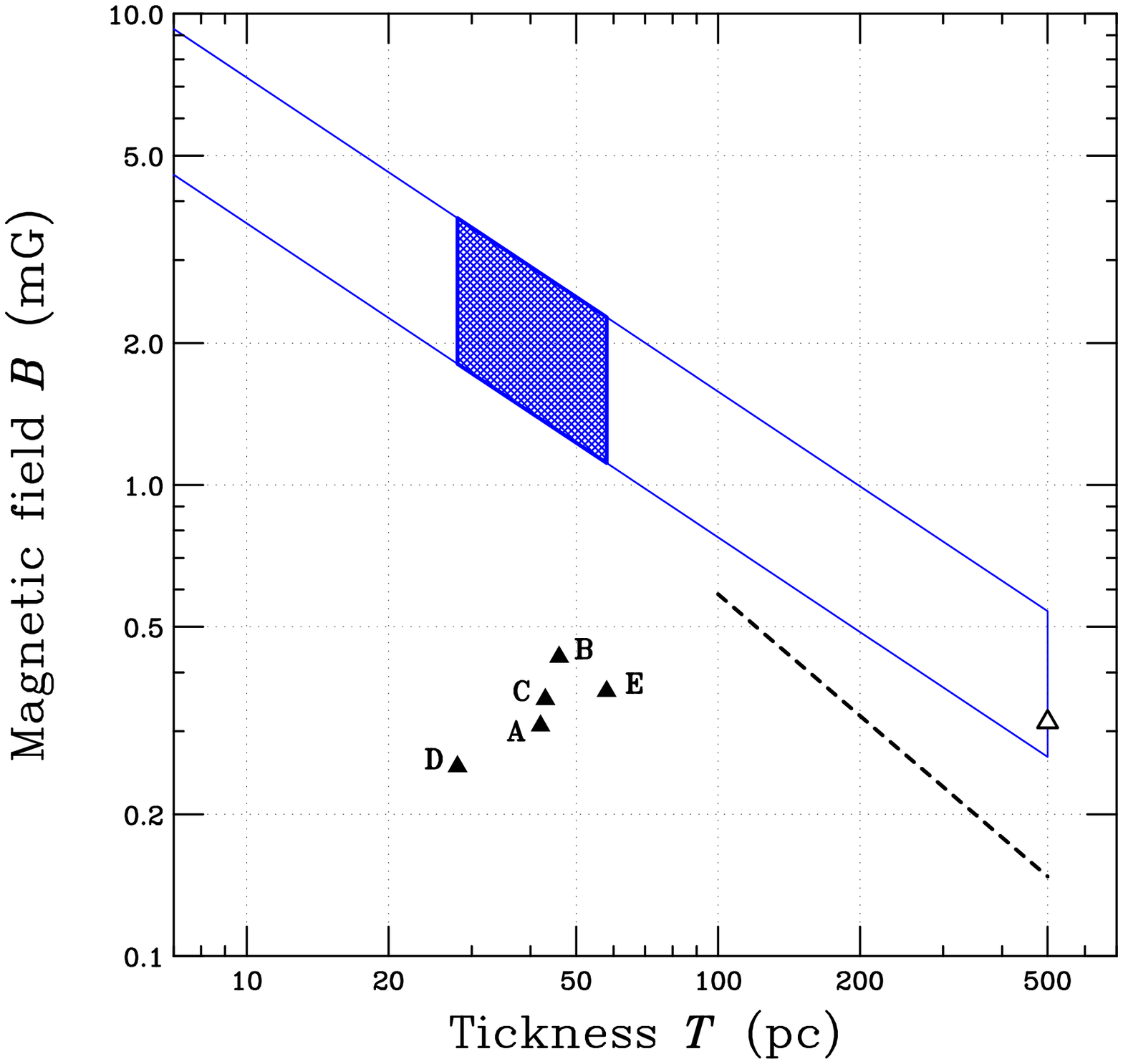}
\caption{
Constraints on the magnetic field $B$ 
of the FIR-to-MIR excess observed from the west hot spot,
as a function of the thickness (or the sphere diameter)
$T$ of the emission region.
The filled and open triangles indicate 
the minimum-energy magnetic field, $B_{\rm me}$,
of the substructures (denoted as A, B, C, D and E) 
and entire west hot spot, respectively. 
The dashed line shows the $B_{\rm me}$ upper limit
for the large source with $T>100$ pc,
estimated from the constraints on the flux density of the excess 
($F_{\nu} < 382$ mJy at 1.67 GHz) 
and on the energy index ($\alpha_{\rm low} < 0.42$).
The magnetic field of the excess,
estimated by inputting the measured break frequency of
$\nu_{\rm b} = 1.6_{-1.0}^{+3.0} \times 10^{12}$ Hz
into Equation (\ref{eq:Bme}), 
is shown with the thin solid line.  
For the estimation, 
a representative downstream flow velocity of $v = 0.3 c$
\citep[][]{kino04} is adopted. 
The upper bound on $T$ corresponds to the diameter 
of the west hot spot \citep[$500$ pc;][]{wilson01}. 
When the excess is assumed to originate in the VLBA substructures
with a thickness of $T=28$--$58$ pc \citep[][]{tingay08},
the magnetic field required by the excess 
is restricted to be within the hatched area.
} 
\label{fig:B-R}
\end{figure}

The area enclosed by the thin solid lines on Figure \ref{fig:B-R} displays 
the acceptable range of the magnetic field strength,
which is estimated from the break frequency of the excess,
$\nu_{\rm b} = 1.6_{-1.0}^{+3.0} \times 10^{12}$ Hz, 
as a function of the region size $T$. 
The flow velocity of $v=0.3c$ is employed for evaluation,
as a representative value of jet-terminal shocks 
in radio galaxies \citep[e.g.,][]{kino04}.
When the FIR-to-MIR excess is interpreted by the substructures 
with a thickness of $T=28$--$58$ pc, 
its magnetic field is limited as $B\simeq 1$--$4$ mG,
as shown with the hatched quadrangle on Figure \ref{fig:B-R}.
If the emission region of the excess is larger than these substructures,
a weaker magnetic field is required (e.g., $B=0.5$--$1$ mG at $T=200$ pc)

With the refined spectral investigation enabled by the SPIRE data, 
the magnetic field strength higher than the minimum-energy value of the substructures,
originally suggested by \citet{isobe17}, is reconfirmed as $B\simeq(3$--$10)B_{\rm me}$. 
Even if the excess is assumed to come from the larger region 
with $T>100$ pc, 
a comparison of the thin solid and dashed lines on Figure \ref{fig:B-R}
indicates that the magnetic field is stronger than $B_{\rm me}$ 
at least by a factor of $\sim 1.5$.
Since the minimum-energy condition is nearly equivalent 
to the energy equipartition between synchrotron-emitting electrons and magnetic field,
the magnetic energy density of the excess is expected to be higher than 
the non-thermal electron one roughly by a factor of $(B/B_{\rm me})^4 = 3^4$--$10^4$ 
(or at least $\sim (1.5)^4$). 

\subsubsection{Possible interpretation} 
\label{sec:B_interpretation}
Because it is difficult to ascribe such a high magnetic dominance to 
a simple plasma compression at the shock alone, 
some other physical processes, 
which selectively enhance the magnetic field, are required. 
Theoretical studies indicate that 
the magnetic field in the downstream region of the shock 
is locally amplified by turbulence 
induced by plasma instabilities and/or inhomogeneities. 
In the case of non-relativistic strong shocks,  
a magnetic-field amplification by more than two orders of magnitude 
is widely predicted by
analytical calculations \citep[e.g.,][]{fraschetti13}
and magnetohydrodynamic (MHD) simulations \citep[e.g.,][]{inoue09,sano12,ji16}. 
A similar level of magnetic-field boost is reported 
even in MHD simulations for relativistic shocks \citep[e.g.,][]{inoue11}. 
This mechanism is invoked to explain the magnetic field 
inferred for X-ray shells/rims \citep[$\sim100$ $\mu$G,][]{vink03,bamba03}
and X-ray hot spots \citep[$\sim1$ mG,][]{uchiyama07} found in supernovae remnants,  
since these magnetic fields are significantly stronger than the interstellar field (a few $\mu$G).
Therefore, it is natural to infer 
that regions with an amplified magnetic field residing in the post-shock medium 
inside the west hot spot are observed as the substructures 
from which the FIR-to-MIR synchrotron excess is radiated. 

\subsection{Electron acceleration process} 
\label{sec:interpretation}
\subsubsection{Doubt as to the standard shock scenario}
It is widely believed that 
the particles are efficiently accelerated by the jet-terminal shock
via the first-order Fermi process 
in the hot spots of radio galaxies \citep{begelman84}. 
It is theoretically shown that this acceleration process
produces a non-thermal electron population with a PL-shaped energy distribution function 
described as $N(\gamma)\propto \gamma^{-p}$ \citep[e.g.,][]{drury83},
in the $\gamma$ range where the radiative cooling is insignificant. 
In the non-relativistic case, 
the spectrum of the accelerated electrons becomes hardest at the strong shock limit 
with the index of $p=2$, 
which corresponds to the synchrotron index of $\alpha=0.5$ (i.e., $p=2\alpha+1$).
For relativistic strong shocks, 
the electron spectrum is suggested to be slightly softer as $p=2.2$--$2.4$ 
\citep[e.g.,][]{kirk00,achterberg01,spitkovsky08}. 
Therefore, the standard strong-shock picture is unable to create 
the flat electron spectrum with $p=1.44\pm0.12$,
required for the 10-pc substructure to reproduce 
the FIR-to-MIR excess with $\alpha_{\rm low} = 0.22 \pm 0.06$.
Even if the 100-pc scale source is considered, 
the spectrum ($\alpha_{\rm low} < 0.42$ and hence  $p<1.84$) is 
still harder than the prediction by the strong shock.
In the following, two alternatives are discussed 
as a viable mechanism to make a hard spectrum with $p<2$ and hence $\alpha<0.5$;
{\it stochastic acceleration via magnetic turbulence} and 
{\it acceleration by magnetic reconnection}.

\subsubsection{Stochastic acceleration} 
\label{sec:stochastic}
Within magnetized plasma flows, 
particles are expected to be powered by magnetic turbulence 
\citep[e.g.,][]{schlickeiser84,howes10}.
The particle acceleration through the magnetic turbulence,
usually called as the stochastic acceleration,
is phenomenologically equivalent to the second-order Fermi acceleration. 
This mechanism was applied to gamma-ray bursts 
\citep[e.g.,][]{asano09, asano15} of which the spectrum 
is conventionally approximated by the so-called Band function \citep{band93},
and to gamma-ray blazars \citep[e.g.,][]{asano18}, 
especially those with an exceptionally hard spectrum \citep{asano14}. 
The stochastic acceleration seems compatible to
the magnetic amplification presented in \S \ref{sec:B_interpretation}. 
It is suggested that 
the synchrotron electrons in the X-ray hot spots of the supernovae remnants, 
where the magnetic field is possibly strengthened by turbulence, 
are energized by the in-situ stochastic acceleration \citep{inoue09}. 

By referring to the simple review by \citet{asano14},
the stochastic acceleration is applied to 
the FIR-to-MIR excess from the west hot spot of Pictor A.
The electron spectrum obtained by the stochastic acceleration
is controlled by the turbulence spectrum. 
In the standard manner, the spectrum of the turbulent magnetic field 
is written as $\delta B^2 \propto k^{-q}$,
where $k$ is the wave number of the turbulence. 
In the steady state, 
the spectrum of the accelerated electrons converges on the PL form as 
$N(\gamma)\propto \gamma^{-(q-1)}$ (i.e., $p = q-1$),
if the effect of electron escape from the acceleration region is neglected. 
The Kolmogorov turbulence \citep{kolmogoro41},
which is often observed in natural phenomena, is represented by $q = 5/3$.
In this case, a "rising" synchrotron spectrum 
with $\alpha = -1/6$ ($p=2/3$) is anticipated,
although such a very hard  spectrum is 
rarely observed from synchrotron objects.
The value of $q>2$ is thought to be unphysical,  
since the acceleration time scale ($\propto \gamma ^{-(q-2)}$) 
becomes shorter for higher energy particles. 
Therefore, the softest spectrum is obtained in the hard-sphere condition with $q=2$,
corresponding to  $p=1$ and $\alpha =0$,
where the acceleration time scale is independent of the electron energy. 
It is reported that the hard-sphere-like turbulence is achieved 
in the presence of energy transfer from small to large scales 
\citep[inverse transfer,][]{brandenburg15}. 

The synchrotron and corresponding electron spectra of the FIR-to-MIR excess,
$\alpha_{\rm low} = 0.22$ and $p=1.44$, respectively
(or $\alpha_{\rm low} < 0.42$ and $p<1.84$ in the case of the 100-pc scale source)
are still slightly softer than the prediction
for the hard-sphere condition ($\alpha=0$ and $p=1$).
In other words, the index of the turbulence required for the excess, 
$q=2.44$ (or $q<2.84$),
appears to be within the unphysical range.
In order for the stochastic acceleration to be acceptable, 
some other processes are necessary to steepen the electron spectrum. 

The above argument neglects, for simplicity, 
the effect of the electron escape from the acceleration region.  
It is confirmed in numerical studies \citep[e.g.,][]{becker06} that 
the spectrum becomes softened by taking the electron escape into account,
since the escape time scale is shorter for higher energy particles. 
Thus, the stochastic acceleration accompanied by the particle escape
is proposed as one of the  plausible mechanisms for the FIR-to-MIR excess.

\subsubsection{Magnetic reconnection} 
\label{sec:reconnection}
Another important process, 
which invokes energy dissipation from the magnetized plasma flow
into high-energy particles,
is provided by magnetic reconnection \citep[e.g.,][]{rowan17}.
The plasma heating and particle acceleration caused by the magnetic reconnection
is actually observed in solar flares \citep[e.g.,][]{tsuneta92}.
There are numbers of application of the magnetic reconnection to
high energy phenomena in various kinds of astrophysical objects, 
including pulsar wind nebulae \citep[e.g.,][]{lyubarsky01},
jets from active galactic nuclei \citep[e.g.,][]{romanova92}, 
and gamma-ray bursts \cite[e.g.,][]{zhang11}.

In order to characterize the magnetized flow, 
the magnetization parameter is widely utilized. 
In theoretical studies of electron-positron pair plasma
\citep[e.g.,][]{guo14,guo15,sironi14,werner16}, 
the magnetization parameter is defined 
as $\sigma = B^2/(4\pi n_{\rm e} m_{\rm e} c^2)$ 
where $n_{\rm e}$ is the number density of electrons and positrons.
This parameter quantifies the relative dominance of the magnetic energy density 
to the electron rest-mass one in the flow. 
The electron acceleration by the magnetic reconnection 
is predicted to effectively work 
for the highly magnetized condition of $\sigma \gg 1$.
The strong magnetic field of $B/B_{\rm me}\simeq(3$--$10)$
estimated for the substructure from the excess 
(or $\gtrsim 1.5$ for the 100-pc scale emission region) 
is reminiscent of the high magnetization. 

One of the remarkable properties of the reconnection acceleration is 
that the process is able to easily construct a hard PL spectrum 
with an electron index of $p<2$.
It is widely predicted by particle-in-cell simulations 
\citep[e.g.,][]{guo14,guo15,sironi14,werner16}
that the hardness of the electron spectrum is clearly dependent on the magnetization; 
the higher $\sigma$ value tends to give the flatter spectral index. 
Based on these simulations,
the spectral index of the FIR-to-MIR excess ($p=1.44$ and $\alpha_{\rm low} = 0.22$)
appears to be achieved in the magnetization of $\sigma = 10$--$50$.
Even though the soft end of the derived index 
($p<1.84$ and $\alpha_{\rm low} <0.42$) is adopted, 
the magnetization parameter of at least $\sigma>5$ is necessary.

A difficulty of the reconnection interpretation is 
found in the highest electron Lorentz factor. 
The numerical studies indicate that
the magnetic reconnection with $\sigma = 10$--$50$  
is possible to accelerate the electrons 
up to the maximum Lorentz factor of  $\gamma_{\rm max} \simeq 100$ \citep{werner16} 
or at most $\lesssim 3000$ \citep{sironi14}. 
These values are highly surpassed 
by the Lorentz factor corresponding to the cutoff frequency of the excess,
$\gamma_{\rm c} = 1.9 \times 10^5 (B/2 {\rm~mG})^{-0.5}
(\nu_{\rm c}/8.8\times10^{13} {\rm ~Hz})^{0.5} $. 
The overall synchrotron spectrum of the excess is unattributable to 
the simple magnetic reconnection alone. 
This problem may get severer,
if the origin of the excess is the 100-pc scale substructures,
since the lower $\gamma_{\rm max}$ value is anticipated for the weaker magnetization.

An interesting idea to overcome this problem is presented in \citet{werner16}. 
In the standard theoretical/numerical studies,
the reconnection is assumed to be induced by cold injection flows. 
If the reconnection is caused by "hot" injection flows 
with an average electron Lorentz factor of $\overline{\gamma}$,
the maximum Lorentz factor is expected to be boosted by a factor of $\overline{\gamma}$
(i.e., $\gamma_{\rm max, hot} = \overline{\gamma}  \gamma_{\rm max}$).
In order to interpret the excess by the reconnection,
the average Lorentz factor of the hot injection flow
is requested to be in the range of $\overline{\gamma}=100$--$1000$ 
($\sim \gamma_{\rm c}/\gamma_{\rm max}$). 
Therefore, the reconnection accompanied by some pre-heating/acceleration mechanisms 
is possible to give a solution to the spectrum of the excess.  

The origin of the excess, regardless of is spatial scale,  
basically resides in the west hot spot. 
Thus, the strong shock, 
which is a source of the main synchrotron component of the west hot spot,
is possible to supply the hot injection flow. 
Another source of the pre-acceleration 
is potentially provided by the substructures themselves. 
It is generally considered in theoretical studies 
\citep[e.g.,][]{quataert1999,lazarian99,lazarian04,zhang11,chael18,chael19} 
that the magnetic turbulence is mediated by the reconnection.
Therefore, it is inferred that 
the reconnection and stochastic accelerations are concurrent in the magnetized plasma. 
If this is the case, the stochastic acceleration in the substructures 
acts as the pre-accelerator.

In summary, the very hard spectrum of the FIR-to-MIR excess observed 
from the west hot spot of Pictor A
is ascribed to the stochastic acceleration and/or magnetic reconnection.
Both processes are thought to have a compatibility to the magnetic amplification 
by the turbulence, which is proposed to explain the strong magnetic field of the excess.
A further support for this idea will be given by 
numerical/analytical evaluation of the acceleration-related parameters,
although it is out of scope of the present study.

\subsection{Implications for the origin of the X-ray emission} 
\label{sec:X-ray}
Finally, the possible mechanism for the X-ray emission and 
its relation to the FIR-to-MIR excess are briefly discussed. 
In spite of the extensive multi-frequency studies of the west hot spot of Pictor A, 
no unified picture has been yet constructed for the origin of its X-ray photons. 
Because of the X-ray slope \citep[$\alpha_{\rm X} = 1.07 \pm 0.11$,][]{wilson01}
significantly steeper than the synchrotron radio one
and of the unusually high X-ray-to-radio flux ratio,
it is difficult to reproduce simultaneously the X-ray and radio-to-optical spectra 
by the standard one-zone SSC model, even if the FIR-to-MIR excess is neglected. 
In order to overcome this problem,
a number of multi-component interpretations were attempted 
\citep[e.g.,][]{wilson01,tingay08}.

The new population of non-thermal 
electrons possibly localized in the substructures within the hot spot 
is identified in the present study 
through the discovery of the FIR-to-MIR synchrotron excess.
However, considering the hardness of the excess 
($\alpha_{\rm low} = 0.22 \pm 0.06$ or conservatively $\alpha_{\rm low} < 0.42$), 
the inverse-Compton emission by this additional electron component is unable to 
give a meaningful solution to the X-ray spectrum,
regardless of its seed photon source.

It is proposed in \citet{tingay08} that 
the X-ray spectrum of the west hot spot of Pictor A is 
created by the substructures via the synchrotron radiation.
They overcame the problem of the X-ray softness 
by assuming that the synchrotron break/cutoff is located 
around the X-ray band.
In order to make the idea of \citet{tingay08} 
compatible to the result of the present study, 
the substructures are requested 
to produce two independent synchrotron electron populations;
one for the FIR-to-MIR excess and the other for the X-ray photons. 
Because these two components 
exhibit a very different spectral shape from each other, 
the acceleration conditions in the individual substructures 
are possible to be heterogeneous. 

When the mG-level magnetic field in the substructures 
required to explain the FIR-to-MIR excess 
is applicable to the X-ray emitting region,  
the synchrotron cooling time scale of the X-ray electrons 
is evaluated as 
$t_{\rm sync} \sim 1$ yr $(B/2~{\rm mG})^{-1.5} (E_{\rm X}/1~{\rm keV})^{-0.5}$ 
where $E_{\rm X}$ is the X-ray synchrotron photon energy. 
This time scale is consistent to the observational fact that 
the hot spot showed an X-ray flux decrease 
by 10\% in $3$ months \citep{hardcastle16}. 
In contrast, the electron cooling time scale 
at the break of the excess is significantly longer as 
$t_{\rm sync} \sim 450$ yr $(B/2~{\rm mG})^{-1.5} 
		     (\nu_{\rm b}/1.6\times10^{12}~{\rm Hz})^{-0.5}$. 
The longer timescale is consistent to the MIR result 
that no significant flux variation is detected 
between the \spitzer\ and \wise\ observations with an interval 
of $\sim 5$ years \citep{werner12,isobe17}.
Thus, the substructure interpretation 
with a magnetic field of a few mG becomes consistent to 
the observed timescale of the X-ray and MIR variations.  
As is implied in \citet{isobe17}, 
the relatively young and elderly substructures are thought to account 
for the X-ray spectrum and FIR-to-MIR excess, respectively. 

Another exotic multi-component idea is presented by \citet{aharonian02};
i.e., proton synchrotron radiation.
They tried to model the X-ray spectrum of extended jet features in several radio galaxies 
by the proton synchrotron radiation.
In their model, the soft X-ray spectrum of the west hot spot of Pictor A   
is interpreted by the effect of the synchrotron cooling or the proton escape. 
In either cases, the proton synchrotron process,
when applied to the west hot spot of Pictor A,
requires a magnetic field of a few mG with a source size of $\sim$ kpc,
which is comparable to the overall size of the hot spot  
\citep[a diameter of 500 pc is commonly adopted in][]{wilson01,tingay08,isobe17}. 
Interestingly, this field strength is comparable to that of the substructures
evaluated by the FIR-to-MIR excess.  
Considering the minimum-energy magnetic field of the entire west hot,
$B_{\rm me} = 310$ $\mu$G
(the open triangle on Figure \ref{fig:B-R}), 
the proton-synchrotron scenario 
inevitably requests the magnetic field amplification 
to take place all over the object.

\subsection{Future perspective} 
\label{sec:future}
The present result clearly demonstrates the great potential of the FIR data 
for investigating the particle acceleration phenomena in the hot spots,
and presumably in other jet-related features.
It is interesting to mine the \herschel\ data archive for other FIR hot spots.
In addition, the next generation space observatory 
Space Infrared Telescope for Cosmology and Astrophysics
\citep[{\itshape SPICA};][]{roelfsema18} 
is expected to make a great progress in this research field. 
These FIR data will give an answer to 
whether the FIR-to-MIR excess and associated physical processes
are a common feature among the hot spots.  

In order to make a definite conclusion to 
the two-zone scenario for the radio-to-optical synchrotron spectrum 
of the west hot spot in Pictor A,  
it is of crucial importance to obtain high-resolution images 
with a sub-arcsec resolution 
between the mm and FIR ranges. 
Such images will be utilized to specify which substructures dominate the excess. 
For this purpose, the Atacama Large Millimeter/submillimeter Array (ALMA) 
is regarded as the ideal instrument. 
In addition, a denser short-baseline coverage enabled by ALMA is 
useful to search for 100-pc scale substructures
which were possibly missed in the VLBA observation by \citet{tingay08}.
Finally, polarimetric information obtained with ALMA is expected 
to be very useful to constrain the magnetic field configuration 
and to infer the particle acceleration process \citep[e.g.,][]{orienti17}.

\acknowledgments
The authors deeply appreciate 
the productive comments from the anonymous referee
to improve the arguments in the present paper. 
The authors are grateful to Dr. K. Asano for his suitable advice 
on the stochastic acceleration. 
The ATCA data of Pictor A were originally provided by Dr. E. Lenc.
This research made use of the data obtained with \herschel,
an ESA space observatory with science instruments provided 
by European-led Principal Investigator consortia and 
with important participation from NASA. 
The present research is supported 
by JSPS KAKENHI Grant Numbers JP18K03656 and JP18H03721.


\end{document}